# Probing thermal magnon current mediated by coherent magnon via nitrogen-vacancy centers in diamond

Dwi Prananto[1*†], Yuta Kainuma[1], Kunitaka Hayashi[1], Norikazu Mizuochi[2], Ken-ichi Uchida[3,4,5],

Toshu An[1‡]

[1]School of Materials Science, Japan Advanced Institute of Science and Technology,

Nomi, Ishikawa 923-1292, Japan

[2]Institute for Chemical Research, Kyoto University, Gokasho, Uji, Kyoto 611-0011, Japan

[3]National Institute for Materials Science, Tsukuba 305-0047, Japan

[4]Institute for Materials Research, Tohoku University, Sendai 980-8577, Japan

[5]Center for Spintronics Research Network, Tohoku University, Sendai 980-8577, Japan

---

[*] e-mail: prananto@jaist.ac.jp

[†] Present address: Materials Science Program, Faculty of Engineering, Niigata University, Niigata City, Niigata 950-2181, Japan

[‡] e-mail: toshuan@jaist.ac.jp

**ABSTRACT:** Currently, thermally excited magnons are being intensively investigated owing to their potential in computing devices and thermoelectric conversion technologies. We report the detection of thermal magnon current propagating in a magnetic insulator yttrium iron garnet under a temperature gradient using a quantum sensor: electron spins associated with nitrogen-vacancy (NV) centers in diamond. Thermal magnon current was observed as modified Rabi oscillation frequencies of NV spins hosted in a beam-shaped bulk diamond that resonantly coupled with coherent magnon propagating over a long distance. Additionally, using a nanodiamond, alteration in NV spin relaxation rates depending on the applied temperature gradient were observed under a non-resonant NV excitation condition. The demonstration of probing thermal magnon current mediated by coherent magnon via NV spin states serves as a basis for creating a device platform hybridizing spin caloritronics and spin qubits.

# I. INTRODUCTION

The utilization of magnons, i.e., the quanta of collective spin excitation, in magnetic media for transmitting and processing information has flourished in the recent decade and is known as magnon spintronics [1–4]. Moreover, the emerging field of spin caloritronics [5], which utilizes the interplay between spin and heat currents, resulted in an alternative strategy in creating more efficient computing devices [6,7] and versatile thermoelectric conversion technologies [8]. The progress in magnon spintronics and spin caloritronics field is benefited from the ubiquitous use of spin transport measurement based on the inverse spin Hall effect (ISHE) [9], in which a paramagnetic heavy metal is patterned on a ferromagnetic medium [2,6,8].

Quantum sensors based on the electron spins in diamond with nitrogen-vacancy (NV) centers have been regarded as eminent sensors for various condensed matter phenomena [10–12], including spin waves, as it offers high spatial resolution at nanoscale enabling to probe fluctuating magnetic fields with broad frequency band from static to GHz, and non-perturbative operation [10,13]. NV centers are well coupled to coherent magnetostatic spin waves (MSWs) owing to their energy matching [14–20]. Recently, magnon population has been measured and controlled via pumping the NV center by spin waves with a single NV spin sensitivity [21–23]. Furthermore, the same effect was observed nonlocally using the ISHE [24]. Additionally, NV spin excitations and modulations via the spin-transfer-torque oscillation of spin waves by electrical methods through the spin Hall effect have been demonstrated recently [25–27].

In contrast, thermally excited magnons with significantly higher energy [28] (defined by $\hbar\omega = k_B T$) than NV spins cannot resonantly excite NV spins, whereas the high-energy magnons can affect the NV relaxation rate in a non-resonant way [29,30]. These high-energy magnon current is known to interact with lower-energy MSWs through the thermal magnon spin-transfer torque [31–36]. Thus, probing thermal magnon current via NV spin can be realized using MSWs as a mediator.

Herein, we report the detection of thermally excited magnon current mediated by MSW by exploiting the thermal magnon spin-transfer torque (Fig. 1), bridging the energy gap between the thermally excited magnons and NV spin. Using an ensemble of NV spins in a bulk diamond, we observed the modification of the magnetostatic surface spin waves' (MSSWs') magnetization dynamics under resonant NV spins excitations influenced by the thermal magnon current in a magnetic insulator yttrium iron garnet (YIG). Besides, under a non-resonant NV spins excitation condition in a nanodiamond, we also observed NV relaxation-rate changes related to the thermal magnon current.

## II. METHODS

We used a liquid-phase-epitaxy grown YIG sample in the form of a trilayer of single-crystalline YIG/gadolinium gallium garnet (GGG)/YIG of thicknesses 100, 550, and 100 μm, respectively, measuring 6 mm × 3 mm (Fig. 2(a)). To improve the lattice matching between YIG and GGG, a small amount of yttrium in the YIG was substituted with bismuth.

Throughout the experiment, external magnetic fields $\pm B_{\text{ext}}$ were applied along the y-axis with a tilted angle $\phi$ to the surface plane of the YIG/GGG/YIG (Fig. 2(a)). Two gold-wire antennas A and B (50 μm in diameter) were overlaid on the surface near both edges of the upper YIG, separated approximately 2 mm away to excite MSSWs by electrical microwave field, and the MSSWs propagates along the $\boldsymbol{k} \parallel \boldsymbol{B}_{\boldsymbol{ext}} \times \hat{\boldsymbol{n}}$ direction ($\hat{\boldsymbol{n}}$ is a vector normal to the YIG's surface) [37]. In this setup, the MSSWs are predominantly excited on the upper YIG layer surface by one of the antennas and propagate to the other end of the sample depending on the polarity of the applied external magnetic field, where $+B_{\text{ext}}$ ($-B_{\text{ext}}$) is along the $+y$ ($-y$) axis (Fig. 2(a) shows the case for antenna A excitation).

We used two types of diamond NV centers: a diamond beam ((110) oriented) measuring 2.5 mm × 0.1 mm × 0.1 mm containing a layer of the NV spin ensemble (occupying a depth ranging up to 70 nm and a mean depth of 40 nm beneath the surface, see supplemental Note 1 [38]), and a nanodiamond with a diameter of approximately 40 nm containing several NV spins (Adámas

Nanotechnologies). It is noteworthy to mention that the use of the diamond beam with a well-known NV axis direction is suitable for efficient resonant NV spin excitations but not for non-resonant excitations owing to the significant distance of approximately 1 $\mu$m separating the NV spins and the YIG surface [20,39]. As shown in Fig. 2(a), the diamond beam was placed on the upper YIG's layer at the middle of its longitudinal direction, where an external magnetic field $+B_{\text{ext}}$ directed along the y-axis ($[00\bar{1}]$ crystal direction of the diamond beam) creates an angle $\phi$ of 32° to the (110) plane (157° to NV3 ($\parallel$ [111])) of the diamond beam (Fig. 2(a)). This setup separates the resonance transitions of the four possible NV spins directing to the $\langle 111 \rangle$ symmetrical axes (NV1 $\parallel [1\bar{1}\bar{1}]$, NV2 $\parallel [\bar{1}1\bar{1}]$, NV3 $\parallel [111]$, and NV4 $\parallel [\bar{1}\bar{1}1]$).

A temperature gradient $\nabla T$ was created along the YIG's longitudinal direction by increasing or lowering the temperature at either site A ($T_A$) or site B ($T_B$). Such temperature control keeps the temperature at the middle of the YIG's longitudinal dimension constant, as well as the diamond beam's temperature, under the application of temperature differences $\Delta T$ up to 10 K (Fig. 2(a)). This was confirmed using the temperature sensing capability of the NV spins [40–42] and infrared thermography (see Supplemental Note 4 and 5 [38]). $\Delta T$ is defined as the difference between $T_A$ and $T_B$ ($\Delta T = T_A - T_B$).

For the optically detected magnetic resonance (ODMR) measurements, the NV spins' ground triplet ($^3A_2$) states, $m_s = 0$ and $m_s = \pm 1$, are optically addressed using an in-house scanning confocal microscope (see Supplemental Note 2 [38]). In this study, spin-state manipulation, $m_s = 0 \leftrightarrow \pm 1$, was performed by the MSSWs-generated electromagnetic microwave radiation (Fig. 2(b)), propagated from one of the gold-wire antennas to the laser spot position separated by approximately 1 mm away [17,18].

## III. RESULTS

### A. Spin wave and NV spin resonance mapping

The MSSWs were excited from antenna A with microwave (MW) power $P_{MW} = 1$ mW in an increasing $+B_{ext}$, and the YIG's global coherent spin-waves resonance spectra were mapped out by performing microwave absorption ($S_{11}$-parameter) measurement using a vector network analyzer (Rohde & Schwartz ZVB8) at $\Delta T = 0$. Figure 2(c) shows a map of the spin-wave spectra, exhibiting lines of resonance of the MSSWs spanning to the higher frequencies from the uniform Kittel mode (ferromagnetic resonance (FMR)). The solid red and yellow lines indicate the NV spins' upper ($m_s = 0 \leftrightarrow +1$) and lower ($m_s = 0 \leftrightarrow -1$) bound resonance transitions defined by the Zeeman energy, respectively [18]. When an energy matching condition between the MSSW and NV spins is fulfilled ($f_{MSW} = f_{NV}$), the NV spins can be coherently excited by the MSSW [17,18]. From the result in Fig. 2(c), we can expect excitations of the NV spins by the MSSWs within the red and yellow lines.

Next, we mapped out the MSSWs-driven NV spins resonance frequencies by performing ODMR spectroscopy with an increasing $+B_{\text{ext}}$ at $\Delta T = 0$ using the diamond beam. Figure 2(d) shows a color map of the MSSWs-driven ODMR in the diamond beam. As expected, only the NV spins' resonance transitions that matched with the MSSW's resonance frequencies underwent a PL intensity quenching as a consequence of the transition from $m_s = 0$ to $m_s = \pm 1$ [17,18]. In Fig. 2(d), only the $m_s = 0 \leftrightarrow -1$ transitions that overlapped with the MSSW's resonance frequencies appeared. Furthermore, by zooming in around the NV3 spectral line in Fig. 2(d), a discretized and broadened resonance line owing to the frequency matching between the NV spins and the MSSWs with different allowed $k$ wavenumbers (Fig. 2(e)) was observed (Fig. 2(f)) [19,43]. The spectra at a matching condition with $+B_{\text{ext}} = 19$ mT, $f_{\text{MW}} = 2.58$ GHz between a MSSW with a specific wavenumber and the NV spins are shown in Figs. 2(g) and (h).

**B. Detection of thermal magnon current via coherent driving of NV spins**

In a magnet under a temperature gradient, thermal magnon current is generated [4,8] and exerts a thermal spin-transfer torque $\boldsymbol{\tau}_{\text{tm}}$ to a precessing magnetization of coherently excited MSSWs (Figs. 1 and 2(b)). The phenomenon has been well known to be detected through microwave response (Yu et al. [36]) and the ISHE [33–35]. Here an ensemble of NV spins in a diamond beam is utilized to detect the thermal magnon current mediated by MSSWs (Figs. 2(a) and (b)). Note that the applied magnetic field is perpendicular to the MSSWs propagation and the temperature gradient direction (Fig. 2(a)), different from that of Yu et al.'s setup [36] (parallel). Within this geometry, thermal magnon current is not detectable through ISHE since there is no measurable ISHE voltage along the paramagnetic heavy metal stripe if it is deposited parallel to spin polarization vector $\boldsymbol{\sigma}$ along the lateral dimension of the YIG as $\boldsymbol{V}_{\text{ISHE}} \parallel \boldsymbol{J}_s \times \boldsymbol{\sigma}$, where $V_{\text{ISHE}}$ and $\boldsymbol{J}_s$ are ISHE voltage and spin current vector, respectively [9,34].

First, the ODMR spectra of the NV spins excited by the MSSW were analyzed under a temperature gradient. We tuned the resonance frequency to one of the matching condition frequencies of 2.58 GHz as shown in Figs. 2(g) and (h) and analyzed the PL contrast of the ODMR as the $\Delta T$ varied. In Figure 3(a), the ODMR spectra with resonance dip at 2.58 GHz ($+B_{\text{ext}} = 19$ mT, $P_{\text{MW}} = 1$ mW) are shown with an increasing $\Delta T$ ($-10$ to $+10$ K). The ODMR's PL contrast is enhanced as $\Delta T$ evolved from positive to negative, though the MSSWs are driven by the same MW power $P_{\text{MW}} = 1$ mW. Their intensities were plotted together with linear fitting in Fig. 3(b). This indicates a change in the amplitude of microwave AC field from the MSSWs [44], as thermal magnon current was generated under the application of temperature gradient in the upper layer of the YIG.

Next, we drove the NV spins into the Rabi oscillations between the $m_s = 0$ and $m_s = -1$ via the MSSW-driven pulse sequence shown in Fig. 3(c) with the same matching condition of 2.58 GHz between the qubit states of $m_s = 0$ and $m_s = -1$ (Fig. 3(d)). The frequency of the Rabi oscillation $\Omega_R^-$ is proportional to the amplitude of the MSSWs oscillating driving field $b_1$ ($\Omega_R^- \propto b_1$). The negative-sign superscript denotes the left-handed polarization component of the oscillating field of the MSSWs driving the NV spins transition ($m_s = 0 \leftrightarrow -1$) [20,45,46]. The Rabi frequency was enhanced for $\Delta T = 0$ to $-10$ K and was suppressed for $\Delta T = 0$ to $+10$ K (Figs. 3(d) and (e)). This is explained by the change of polarity of the thermal magnon spin-transfer torque [36] (Fig. 1). The amplitude of the Rabi field $b_R^-$, defined as an effective oscillating electromagnetic field acting at the NV position above the YIG surface (Fig. 2(b)), can be estimated from the Rabi frequency through the relation $b_R^- = \Omega_R^-/\gamma_e$ [14,45,46], with $\gamma_e = 2\pi \cdot 28$ GHz/T being the gyromagnetic ratio of electrons. The Rabi field amplitude $b_R^-$ evolved from $19 \pm 0.5$ μT at $\Delta T = 10$ K to $26 \pm 0.4$ μT at $\Delta T = -10$ K, based on its plot as a function of $\Delta T$ (Fig. 3(f)), indicating a change of approximately $18 \pm 1$ % from $22 \pm 0.6$ μT at $\Delta T = 0$.

The unidirectional propagation of the MSSWs is inverted according to $\mathbf{k} \parallel \mathbf{B}_{\text{ext}} \times \hat{\mathbf{n}}$ [47,48] by applying different polarity of $B_{\text{ext}}$ at the upper YIG surface and in this condition the thermal spin-transfer torque is applied with different polarity [36]. Hence, we can expect to observe the same but inverted sign effect when we switch the external magnetic field to the $-y$ axis (assigned as $-B_{\text{ext}}$) and launch the MSSWs from the antenna B [36]. We tuned the NV resonance frequency to a matching condition of 2.60 GHz ($-B_{\text{ext}} = 19$ mT, $P_{\text{MW}} = 1$ mW). As expected, the Rabi frequency was suppressed for $\Delta T = 0$ to $-10$ K and was enhanced for $\Delta T = 0$ to $+10$ K (Figs. 3(g) and (h)). In this geometry, the Rabi field amplitude is estimated to evolve from approximately $18 \pm 0.6$ µT at $\Delta T = -10$ K to approximately $22 \pm 0.3$ µT at $\Delta T = +10$ K (Fig. 3(i)), indicating a change of approximately $16 \pm 2$ % from $19 \pm 0.5$ µT at $\Delta T = 0$.

The observed effect can be interpreted as a thermal magnon spin-transfer torque $\tau_{tm}$ via the thermal magnon current generated by a temperature gradient [4,49,50], which interacts with the MSSW and relaxes by transferring its spin angular momentum (Fig. 1). The transfer of spin angular momentum contributes to the development of the thermal magnon torque $\tau_{tm}$, which alters the MSSW's magnetization dynamics [31,32,36] and perceived by the NV spins as an altering Rabi field amplitude (Fig. 2(b) and see Supplemental Note 9 [38]):

$$b_R^- \propto \lambda M_s \frac{\gamma_e b_{\text{MW}}}{(\alpha_i + a\nabla T)\omega_r} k e^{-kx}, \tag{1}$$

where $\lambda$, $M_s$, $b_{\text{MW}}$, $\alpha_i$, $\omega_r$, and $x$ are respectively the proportionality constant, saturation magnetization of the YIG, microwave field driving the MSSWs, intrinsic damping parameter of the YIG, resonance frequency of the MSSW, and the distance separating the NV spin and the magnetization precession. The contribution from the thermal magnons can be quantified by the thermal magnon damping parameter, which is proportional to the temperature gradient $\alpha_{\text{tm}} = a\nabla T$ (see Supplemental Note 8 [38]). Using a constant $a$ in Eq. (1) as a fitting parameter, $\alpha_{\text{tm}}$ was estimated to be $(10 \pm 0.9) \times 10^{-4}$ for $+B_{\text{ext}}$ and $(4.3 \pm 1) \times 10^{-4}$ for $-B_{\text{ext}}$ using an effective temperature difference of $\Delta T_{\text{eff}} = 6.6$ K over 2 mm distance at the YIG's top surface under an applied $\Delta T = 10$ K.

The thermal magnon damping parameter values agree well with those reported previously [33–36], confirming the existence and contribution of thermal magnon current in the evolution of MSSW magnetization dynamics [26,31,32,36]. Furthermore, we confirmed our observation of the thermally excited magnon current electrically by analyzing the spin-wave resonance linewidth from the absorption microwave signal ($S_{11}$) (see Supplemental Note 7 for the experimental details and data [38]).

### C. Local detection and non-resonant NV spin excitation

We extended the capability to detect the thermally excited magnon current locally and non-resonantly to NV spin transition frequency via a small number of NV spins in a nanodiamond (Fig. 4(a)). The nanodiamonds with 40 nm of averaged diameter were transferred to the middle of the YIG's longitudinal direction by dropping a small amount of nanodiamond solution with a micropipette.

With the same setup and technique as in the experiment using the diamond beam, we mapped out the ODMR spectra of the NV spins in a nanodiamond to obtain information regarding the coupling between the long-distance propagating magnons and the NV spins. Figure 4(b) shows the magnon-driven ODMR spectral map exhibiting PL quenching at the resonance transition ($m_s = 0 \leftrightarrow -1$) of the NV spins together with PL image of the nanodiamond used in the measurement (inset) ($P_{MW} = 1$ W). Additionally, a strong non-resonant PL quenching was observed away from the NV spin transitions [15,39] at frequencies ranging from 2.5 to 2.7 GHz at the $+B_{ext}$ between 11.5 and 13.5 mT (Fig. 4(b)), where the MSSWs with higher $k$ wavenumbers are within a range as observed in Fig. 2(e).

Next, we performed longitudinal spin relaxation measurements, in which the NV spins were polarized to $m_s = 0$ by the first laser pulse, followed by a dark time $\tau$ before another laser pulse was applied to read the remaining population (Fig. 4(c)). By varying $\tau$, the time-trace relaxation of the $m_s = 0$ state to its equilibrium state was observed. Under the application of $\nabla T$ to the YIG, MW pulse with the frequency of 2.66 GHz and $+B_{ext}$ and $-B_{ext} = 13$ mT (marked by dashed-black circle in Fig. 4(b)) was applied with $P_{MW} = 1$ W during time $\tau$.

Figures 4(d) and (e) show the measurements of NV spin longitudinal relaxation rate $\Gamma$ as a function of applied temperature gradient with MW drive in the YIG under opposite polarity of external magnetic fields, $+B_{ext}$ and $-B_{ext}$. For $+B_{ext}$, the longitudinal relaxation rate increased for $\Delta T = 0$ to $-10$ K and decreased for $\Delta T = 0$ to $+10$ K (Fig. 4(d)). Opposite polarity of slope-change of $\Gamma$ was observed when the polarity of $B_{ext}$ is inverted (Fig. 4(e)), reasonable with the MSSW's unidirectional propagation character.

Here, we assume that the observed effect is originated from the modulation of magnon density at NV-resonant frequency via the scattering between the non-resonant MW-excited magnons and the thermal magnons [21,29,31,32]. In this case, $\Gamma$ is related to the oscillating AC magnetic field amplitude generated by the NV-resonant magnons, as described by $\Gamma \sim \frac{\gamma_e}{2}|B_\perp|^2$, with $|B_\perp|^2$ is the AC magnetic field component perpendicular to the NV's quantum axis [14]. By assuming that the AC magnetic field from the NV-resonant magnons evolved proportionally with the increase or decrease of magnetization precession of the MW-excited magnons [21] and based on the fact that the magnetization precession evolved under a variation of $\nabla T$ (Equation (1) and Figs. 3(f) and (i)), we can approximate an equation relating the longitudinal relaxation rate $\Gamma$ and temperature gradient as [14,19,20,26]

$$\Gamma \propto \frac{\gamma_e^2}{2} \left| \frac{\lambda M_s \gamma_e b_{MW}}{(\alpha_i + a\nabla T)\omega_r} k e^{-kx} \right|^2. \tag{2}$$

The data in Figs. 4(d) and (e) were fitted with equation (2), and $\alpha_{tm}$ was estimated as $(4.3 \pm 1) \times 10^{-4}$ for $+B_{\text{ext}}$ and $(2.5 \pm 0.9) \times 10^{-4}$ for $-B_{\text{ext}}$, that show a good agreement with those estimated from the Rabi oscillation experiments. We note that the temperature measurements at the middle of YIG using bulk diamond beam and infrared thermography (see Supplemental Note 4 and Supplemental Note 5 [38]) confirmed a base temperature change of less than 1.5 K, which will give 0.7 % of the change in $\Gamma$ [21]. This change of $\Gamma$ is small compared with the observed change of about 37.5 % (for $+B_{\text{ext}}$) and 23 % (for $-B_{\text{ext}}$) under the applied $\Delta T$ from $+10$ K to $-10$ K to the YIG, showing that the observed effect is not due to the base temperature change in the nanodiamond.

## IV. DISCUSSION

We demonstrated the detection of thermally excited magnon currents mediated by MSSWs via NV spins, where the thermal magnon spin-transfer torque emanated from the thermal magnon current altered the MSSWs magnetization precession when the YIG sample was subjected to a temperature gradient. The modulation of the magnetization dynamics of the MSSWs was perceived by the NV spins as the alteration of the Rabi oscillation frequency with the resonant NV spin excitation using a diamond beam.

Besides, the longitudinal spin-relaxation rate change was observed with a non-resonant NV spins excitation using a nanodiamond. The possible explanation for the observed effect at non-resonant excitation may come from the four-magnon scattering process, where a magnon at the microwave frequency scatters with a thermal magnon resulting in two additional magnons, one of which possesses a frequency resonates to the NV frequency [29,30]. The increase or decrease in the relaxation rate as a function of a temperature gradient indicates the modulation in the population of the thermal magnon [Figs. 4(d) and (e)]. However, to nail down a definite mechanism, it will require further experiments through changing excitation parameters, and also using a nanodiamond or diamond nanobeam with a well-defined NV axis [21,39].

This study provides a detection tool for thermal magnon currents via NV centers, which can be located locally and in a broad range of distances to spin waves. This feature cannot be obtained if only conventional methods, such as ISHE, are used to investigate magnon dynamics, as the conventional method requires a relatively large electrode and specific configurations with proximal distance to the spin waves. Owing to the NV spin's single spin detection sensitivity enabled by its atomic-scale size [51], nanoscale probing and imaging of thermal magnon dynamics can be realized in the future. For example, a scanning probe-based NV magnetometry [13] will be useful for studying the nonuniformity of the thermal magnon current throughout the material at the nanoscale. Such a measurement will be impractical through patterning a large area of a paramagnetic metal for ISHE measurements. A study of the thermal magnon dynamics with high spatial resolution can provide insights into practical applications in spin caloritronics and magnon spintronics [14,19,25].

## ACKNOWLEDGMENTS

We thank E. Abe for fruitful discussions. This study was supported in part by JSPS KAKENHI (18H01868, 18H04289, and 19K15444), Japan, by JST CREST (JPMJCR1875 and JPMJCR17I1), and by JST A-STEP (JPMJTM19AV), Japan. N.M. acknowledges support from KAKENHI (15H05868), MEXT Q-LEAP (JPMXS0118067395), Japan.

**FIGURES AND CAPTIONS**

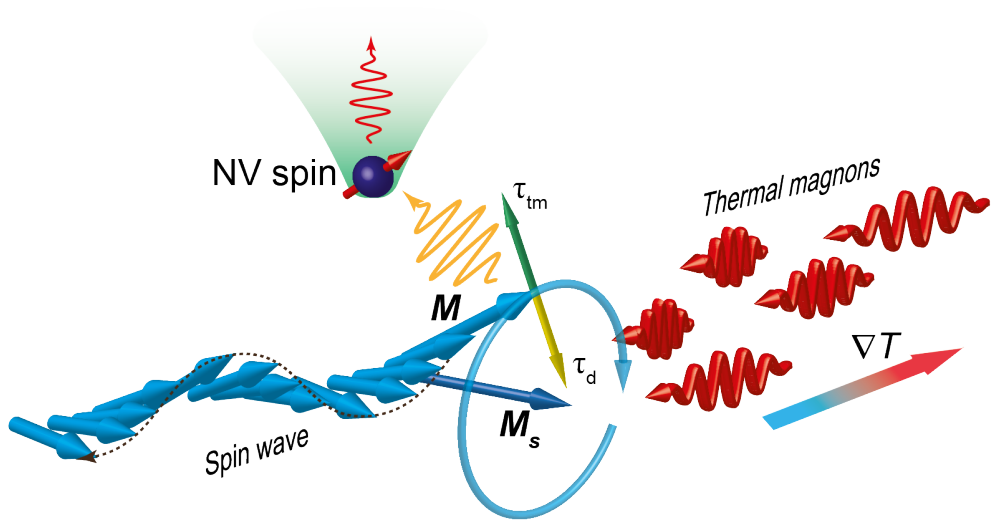

FIG. 1. Mechanism of thermal magnon current detection via NV center. In a magnet, thermal magnon current is created by applying temperature gradient $\nabla T$, exerting a torque $\boldsymbol{\tau_{tm}}$ (with the damping torque $\boldsymbol{\tau_d}$) to spin wave (coherent magnon)'s precessing magnetization $\boldsymbol{M}$ excited by a microwave AC field. Then, information of the thermal magnon current can be probed by nitrogen-vacancy (NV) center spin in diamond near the magnet through the spin wave.

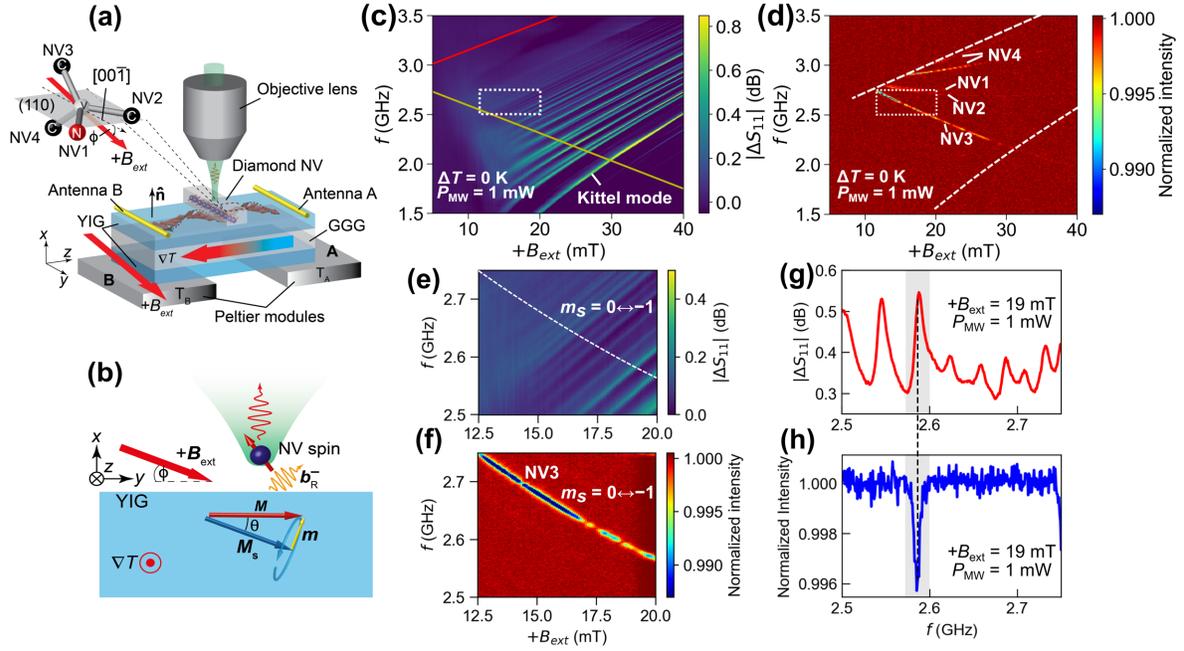

FIG. 2. Experimental setup with a diamond beam and mapping of spin waves and NV spin resonance spectra. (a) Experimental setup for probing thermally excited magnon current via NV centers in a diamond beam centered on the upper YIG's surface. (b) A schematic toy model of thermal magnon current spin-transfer torque system with an NV spin (dark blue ball with red arrow) and a precessing magnetization $\boldsymbol{M}$ in YIG (red arrow) under a magnetic field $\boldsymbol{B}_{\text{ext}}$. The transverse component $\boldsymbol{m}$ of $\boldsymbol{M}$ produces an AC magnetic field amplitude $\boldsymbol{b}_R^-$ that drives the NV spin into its Rabi oscillation. Thermal magnon spin-transfer torque produced by the temperature gradient $\nabla T$ is exerted to the $\boldsymbol{M}$ resulted in the modification of $\boldsymbol{b}_R^-$. $\boldsymbol{M}_s$ is the saturation magnetization of the YIG. (c) Microwave absorption ($P_{\text{MW}} = 1$ mW) spin-wave

resonance spectra as a function of externally applied magnetic field $+B_{ext}$ (Solid lines indicate the upper (red, $m_s = 0 \leftrightarrow +1$) and the lower (yellow, $m_s = 0 \leftrightarrow -1$) bounds of possible ground state resonance transition of NV spins). MSSWs are observed at higher frequencies above the Kittel mode (FMR). (d) ODMR spectra of the NV spins in the diamond beam as a function $+B_{ext}$ field. The region between the two dashed white lines indicates the resonance frequency band of MSSWs. NV1 to NV4 indicate the four possible NV spins directed to $\langle 111 \rangle$ symmetrical axes as shown in (a). (e) Zoomed spin-wave resonance spectra at dotted-white square in (c). Dashed-white line indicates resonance transitions of NV3 spins in (d). (f) Zoomed ODMR spectra in dotted-white square in (d), showing a discretized ODMR resonance line of NV3 owing to the crossing with MSSWs' resonant frequencies. (g) Line cut of (e) at $+B_{ext} = 19$ mT. (h) Line cut of (f) at $+B_{ext} = 19$ mT. Both (g) and (h) show a matching condition at frequency of 2.58 GHz.

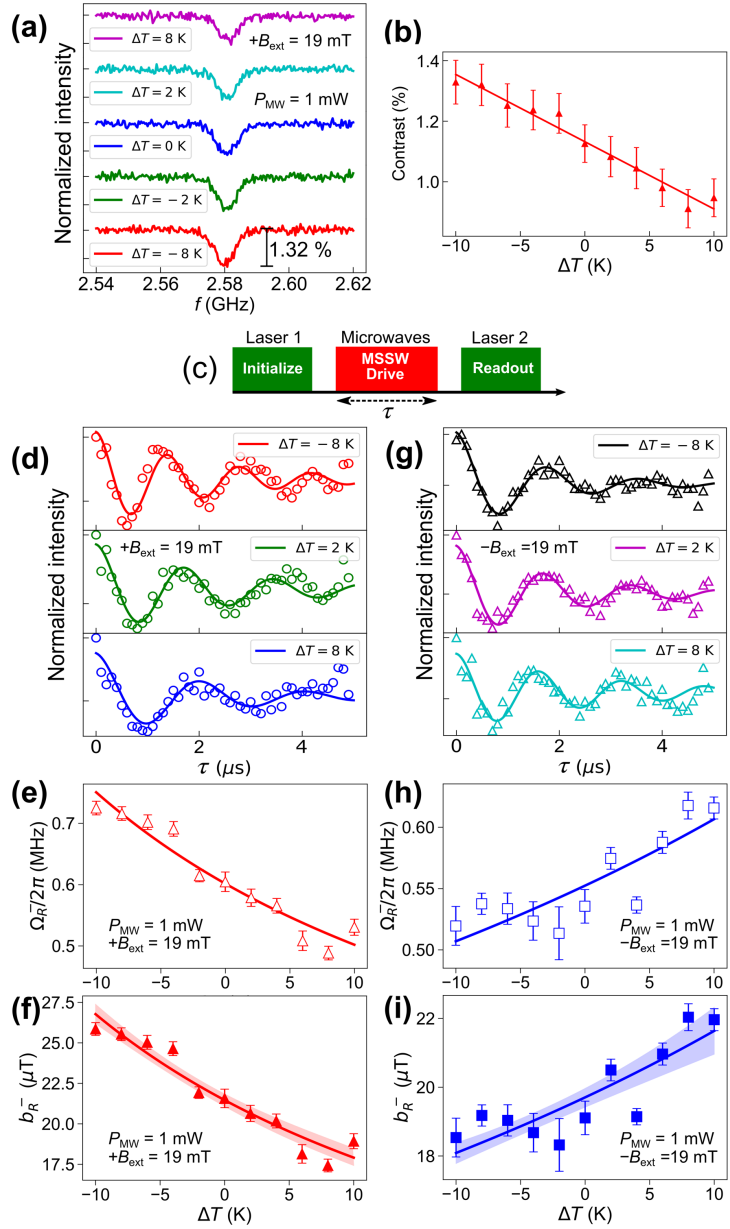

FIG. 3. ODMR spectra and Rabi oscillation frequencies under temperature differences. (a) ODMR spectra with various temperature differences $\Delta T$ applied to the YIG at $+B_{\text{ext}} = 19$ mT and $P_{\text{MW}} = 1$ mW. ODMR dip contrast evolved monotonically (solid line) with $\Delta T$ applied to the YIG. (b) ODMR contrast as a function of $\Delta T$ applied to the YIG. The error bars were obtained from the standard deviation error of the curve fitting to the data in (a) using a single Lorentzian function (not shown here). (c) Measurement protocol to excite Rabi oscillation on the NV spins. Laser pulse 1 initializes the NV spins to the $m_s = 0$ state followed by a microwaves pulse with duration $\tau$ that drives MSSW in the YIG which then excite the NV spins to the $m_s = -1$ state. Laser pulse 2 probes the remaining NV spins population at the $m_s = 0$ state. $\tau$ is varied to produce a stroboscopic oscillation between $m_s = 0$ and $m_s = -1$ states. (d) Rabi oscillations at three different $\Delta T$ applied to the YIG for $+B_{\text{ext}} = 19$ mT. Frequency of Rabi oscillation evolved with applied $\Delta T$. Colored solid lines are damped sinusoidal function. (e) Variation in Rabi oscillation frequency $\Omega_R^-/2\pi$ with $\Delta T$. (f) Calculated Rabi field amplitude $b_R^-$ inferred from the Rabi frequency in (e) as a function of applied $\Delta T$. (g)-(i) Rabi oscillations, Rabi frequencies, and Rabi field amplitudes, respectively as a function of $\Delta T$ with $-\boldsymbol{B}_{\text{ext}} = 19$ mT. The error bars in (e), (f), (h), and (i) were obtained from the standard deviation of the curve fitting to the data in (d) and (g), respectively, with a damped sinusoidal function. Colored solid lines in (e), (f), (h), and (i) are fittings to equation (1). Shaded red and blue areas in (f) and (i) are the possible variation in the fitting curve based on the uncertainties of the fitting parameters in equation (1) (see Supplemental Note 10 [38]).

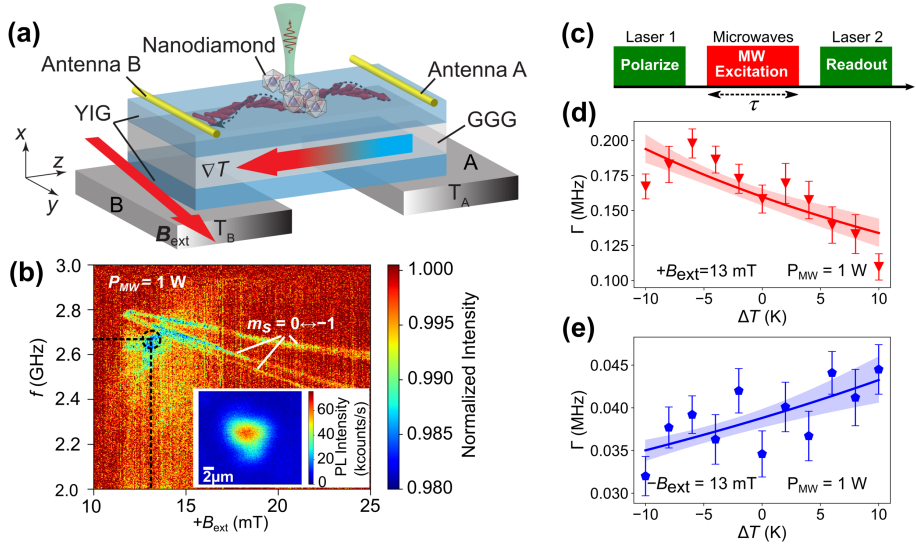

FIG. 4. Local detection of thermal magnon current using a nanodiamond. (a) Experimental setup for local detection of the thermally excited magnon current with a nanodiamond containing several NV spins. (b) ODMR spectra map of the NV spins in the nanocrystal diamond on the YIG under zero temperature difference $\Delta T$ ($P_{MW} = 1$ mW). Non-resonant PL quenching was observed beside the straight lines of the NV spins' resonant transitions $m_s = 0 \leftrightarrow -1$. Inset shows a fluorescence image of the nanodiamond used in the measurement. (c) Longitudinal spin relaxation rate measurement protocol to detect the thermally excited magnon current comprising of a polarizing laser pulse followed by a variable duration $\tau$ pulse of the non-resonant NV spin excitation via MSSW at the frequency of 2.66 GHz with $\pm B_{ext} = 13$ mT (dashed black circle in (b)). (d), (e) NV spin relaxation rate $\Gamma$ as a function of $\Delta T$ applied to the YIG for $+B_{ext}$ ((d)) and $-B_{ext}$ ((e)). Solid red line in (d) and blue line in (e) are the fitting curve to the data with a model in equation (2). The error bars in (d) and (e) were obtained from the standard deviation of the curve fitting to the longitudinal spin relaxation data with a single exponential function (see Supplemental Note 6 [38]). Shaded red and blue areas in (d) and (e) are the possible variation in the fitting curve based on the uncertainties of the fitting parameters in equation (2) (see Supplemental Note 10 [38]).